# Active metasurfaces: lighting the path to commercial success

Tian Gu[1,2,*], Hyun Jung Kim[3,4,*], Clara Rivero-Baleine[5,*], and Juejun Hu[1,2,*]

[1]*Department of Materials Science & Engineering, Massachusetts Institute of Technology, Cambridge, Massachusetts, USA*
[2]*Materials Research Laboratory, Massachusetts Institute of Technology, Cambridge, Massachusetts, USA*
[3]*National Institute of Aerospace, Hampton, Virginia, USA*
[4]*NASA Langley Research Center, Hampton, Virginia, USA*
[5]*Missiles and Fire Control, Lockheed Martin Corporation, Orlando, Florida, USA*

*[*]*gutian@mit.edu, hyunjung.kim@nasa.gov, clara.rivero-baleine@lmco.com, hujuejun@mit.edu*

**Abstract**

Active optical metasurfaces are rapidly emerging as a major frontier in photonics research, development, and commercialization. They promise compact, light-weight, and energy-efficient reconfigurable optical systems with unprecedented performance and functions that can be dynamically defined on-demand. Compared to their passive counterparts, the reconfiguration capacity of active metasurfaces also set additional challenges in scalable design, manufacturing, and control toward their practical deployment. This perspective aims to review the state-of-the-art of active metasurface technologies and their applications while highlighting key research advances essential to enabling their transition from laboratory curiosity to commercial reality.

**Main**

Optical metasurfaces are artificial media comprising planar arrays of sub-wavelength structures (commonly called meta-atoms). With their now well-recognized advantages in optical performances, form factor and cost, metasurfaces are witnessing a move toward commercial adoption: a solid lineup of large corporations as well as a cohort of aspiring start-up companies are heavily investing on R&D in this field.

Active metasurfaces introduce a new dimension to the space, enabling dynamic tuning of optical functions while promising wide-ranging applications in analog computing[1], data communications[2], optical camouflage[3], reconfigurable imaging[4], light detection and ranging (LiDAR)[5], display[6], imaging spectroscopy[7], nonreciprocal photonics[8], and many others. As the interest in active metasurfaces percolates from academia to industry, important questions arise regarding when and how their transition from lab to market will flourish. To address such questions, this perspective provides a bird's eye view on current state-of-the-art of active optical metasurface technologies, surveys emerging applications capitalizing on their unique attributes, and scrutinizes the technological gaps that need to be filled to transform the prospective applications into reality.

**The present state-of-the-art in active metasurface technologies**

Active tuning schemes of metasurfaces can in general be classified into two categories, one where the optical responses of the meta-atoms are modified and the other relying on mechanical movement of meta-atoms. The former scheme usually involves modulating the optical properties of the meta-atoms or their surrounding material, for instance via free carrier injection[9], Pockels effect[10], quantum confined Stark effect[11], thermo-optic coupling[12], electrochromism[13], magneto-optical interaction[14], and structural transitions in various materials[15–17], whereas the latter can leverage either macroscopic displacement[18]/deformation[19] or micro-electromechanical system (MEMS) actuation[20]. State-of-the-art performances of these tuning mechanisms as well as their respective fundamental limits in optical loss, speed, and endurance are summarized in Table 1.

Besides key performance attributes defined by optical contrast, loss, speed, endurance and volatility, the likelihood of an active metasurface technology to enter mainstream adoption in the near- or mid-term is largely dictated by its technology and manufacturing readiness levels (TRL and MRL). From this technology maturity perspective, liquid crystal (LC) and MEMS based metasurfaces are among the most established candidates, as both make use of proven industry-standard technologies to introduce active tuning capabilities and can also take advantage of an established industrial ecosystem to facilitate high-volume device manufacturing and packaging. Other promising contenders involve new materials such as transparent conducting oxides (TCOs) and chalcogenide phase change materials (PCMs). Even though these materials are not part of the standard offerings of most silicon foundries today, they are readily amenable to foundry-compatible backend integration: TCOs constitute an integral element of modern-day display panel production process and PCMs have already become a key ingredient in commercial nonvolatile memories. Integration of these new materials and metasurface architectures into mainstream manufacturing processes will be motivated by practical application demands as discussed in the succeeding section.

**Application prospects: where they will shine**

Before delving into the applications of active metasurfaces, one should be reminded that actively tunable optics is not a new concept. Technologies exemplified by adaptive optics harnessing deformable mirrors for real-time wavefront correction in astronomical telescopes and spatial light modulators (SLMs) building on digital light processing or liquid crystal on silicon (LCoS) have been meticulously perfected over the past decades. So, what do active metasurfaces have to offer?

Their drastically reduced Size, Weight and Power (SWaP) characteristics promise reconfigurable optical systems that are ultracompact, light-weight, energy-efficient and rugged. Their optically-thin, pixelated device architecture further enables fast tuning mechanisms not compatible with conventional bulk optics. Space applications represents one of the emerging arenas where these characteristics are highly prized (Box 1). Moreover, such size down-scaling does not come with the usual penalties of compromised optical quality or lack of fine control, thereby setting active metasurfaces apart from other competing tunable micro-optics technologies relying on electrowetting, liquid metals, and soft elastomeric optics. Active metasurfaces are thus also well poised for applications such as augmented/virtual reality (AR/VR) and point-of-care or minimally invasive biomedical imaging, where form factor and optical precision are equally critical.

Another defining character of active metasurfaces is their capacity for on-demand wavefront manipulation down to the sub-wavelength scale. This unprecedented granularity permits light bending at extreme angles that traditional refractive or diffractive optics cannot accommodate while effectively suppressing spurious diffraction orders[21–23]. Beam steering devices with high efficiency and large field-of-view (FOV) can be created capitalizing on this feature (Box 2). The agile beam control capability, coupled with high spatial density afforded by metasurface optics, potentially envisions glasses-free 3-D displays offering high resolution, large viewing angle, and full color coverage (Box 3). Finally, the ability to engineer a metasurface's phase profile in an almost arbitrary manner proves valuable to computational imaging and sensing, since the transfer function of the frontend meta-optics can be co-designed holistically with the backend processing algorithm to enhance signal extraction and maximize the signal-to-noise ratio[24]. It has been shown that active meta-optics designed using such an approach can yield *optimal* imaging systems capable of multi-dimensional (spatial, spectral, polarization, light field, etc.) information retrieval[25].

Finally, since many active metasurfaces are produced using semiconductor nanofabrication technologies, they can be seamlessly integrated with semiconductor electronic and photonic devices to create 'metasurface-augmented' optoelectronics with novel functionalities, whereas such wafer-level integration is often challenging or impractical for conventional tunable micro-optics.

These unique advantages presented by active metasurfaces foreshadow an array of potential applications outlined in Table 2. The key takeaway message is that there is no "one-size-fits-all" solution, because each application prioritizes a different set of performance metrics that none of the active metasurface technologies today (Table 1) can simultaneously meet. Table 2 also distinguishes two types of tuning schemes: discrete and continuous. In the former case, the metasurface only accesses a small number of optical states, and such discrete tuning over multiple arbitrary phase profiles can be accomplished by collective switching of all meta-atoms across the aperture[26]. On the other hand, it is generally believed that continuous tuning, where a large number or a continuum of states are mandated, necessitates independent control of individual or small groups of meta-atoms. In the following section, we will propose a new concept defying this conventional wisdom to achieve continuous tuning with a much simplified switching fabric. Other research challenges that need to be addressed to fulfill the application demands will also be elaborated.

**Box 1. Active metasurfaces in aerospace applications**

The growth in aerospace systems has been underpinned by increasing capabilities being packed into smaller and lighter spacecrafts, which requires robust, light-weight components to deliver enhanced science data products while constrained by lean SWaP budgets. Future growth in the capabilities of earth observation, deep space, and planetary surface missions using miniaturized spacecraft platforms can only be sustained by innovations in the design of remote sensors and other sub-systems. Enhanced tunability and reconfigurability empowered by active metasurface optics will be game changers for aerospace remote sensing applications, from LiDAR to new imagers, detectors and sensors. Active metasurface topics of interest to aerospace remote sensing include:
- ✓ beam shaping for antennae (optical and microwave wavelengths),
- ✓ reconfigurable (real-time phase-corrective) lenses and planar adaptive optics for imaging, optical communications, and high-gain antennae,
- ✓ beam steering for radar and LiDAR scanning systems, flat panels, and mobile communication antennae (optical, microwaves and millimeter-wave wavelengths), and
- ✓ tunable filter and spatial light modulators for imaging spectroscopy.

As an example, exoplanet imaging in space requires real-time wavefront corrections to mitigate the effects of thermal gradients, optical imperfections, and diffraction issues. An active exoplanet imaging mission at present is the James Webb Space Telescope, which is equipped with a sophisticated NIRCam having several coronagraphs for imaging in the infrared. Active metasurface optics could lead to a space-based correction system with major SWaP advantages. This would permit characterization of the light from an exoplanet and even allow direct imaging of the planet via active cancelation of high frequency spatial and temporal aberrations.

Another application that active metasurfaces are poised to transform is multispectral imaging. Recently, PCM-integrated tunable optical filters have shown promise as multifunctional wide-band replacements for bulky filter wheels in spaceborne remote sensing sub-systems[7,27,28]. This is achieved through the integration of PCM into a plasmonic nanohole metasurface to effectively tune the transmission passband of optical filters in real time. The solid-state metasurface filter is fast, have no moving parts, and does not require continuous input power to maintain the filter characteristics. Their applications include astronaut health monitoring, atmospheric gas sensing, and space launch vehicle thermal imaging as summarized conceptually in the figure.

As a specific use scenario, consider a tunable filter integrated remote temperature measurement system to collect calibrated multispectral images of air/spacecraft during ascent in order to validate the thermal protection systems of the vehicles. To make these measurements, state-of-the-art systems rely on motorized filter wheels which hold several (typically ~ 5) single-notch optical filters and rotate between filters to take new spectral measurements. The filter wheels offer no real-time tunability and are limited in overall bandwidth and temporal resolution, therefore missing out on important spectral and temporal data. In contrast, PCM-based active metasurface filters can be tuned within microseconds, enabling real-time thermography and imaging spectroscopy with high data throughput. Furthermore, LiDAR missions utilize the same filter wheels for chemical remote sensing. The broad waveband and robust tuning applicability of the PCM-based tunable filter provide unprecedented views of earth's atmospheric constituents and surface altimetry, significantly advancing science objectives across multiple disciplines. These features, in addition to the orders-of-magnitude reduction in total SWaP, envisage that spaceborne systems incorporating PCM-based actively tunable filters will continue to be in demand into the foreseeable future.

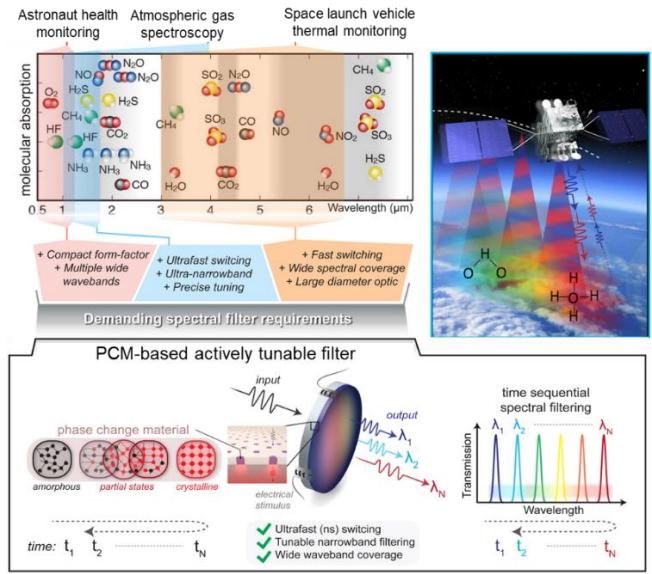

**(Top-left) Characteristic absorption bands of molecules of interest to aerospace applications as well as application-specific requirements on the tunable spectral filter; (top-right) satellite-based multispectral imaging for monitoring of atmospheric gas species; (bottom) operating principle of the PCM-based metasurface filter: the PCM's structure can be continuously tuned in a non-volatile manner via voltage pulses, thereby spectrally shifting the passband of the metasurface.**

### Box 2. Active metasurfaces for beam steering

Optical beam steering devices are gaining importance by the day with prospective applications in LiDAR for autonomous vehicles, remote sensing, and displays in AR/VR modules. A common embodiment of metasurface-based beam steering is to use the metasurface as an optical phased array (OPA). For normal incident input beam with wavelength $\lambda$, the deflection angle $\theta$ of the beam exiting from the metasurface OPA is given by:

$$\frac{2\pi}{\lambda} \cdot \sin\theta = \frac{\Delta\varphi \pm m \cdot 2\pi}{\Lambda}$$

where $\Lambda$ is the meta-atom pitch, $\Delta\varphi$ gives the phase delay between two neighboring meta-atoms ($0 \leq \Delta\varphi < 2\pi$), and $m$ denotes the aliasing order. Ideally, the equation should only be satisfied when $m = 0$, thereby yielding a single solution of $\theta$ to eliminate aliasing. In an OPA with a FOV covering $-\theta_{max}$ to $\theta_{max}$, this condition becomes:

$$\Lambda = \frac{\lambda}{2\sin\theta_{max}}$$

to ensure solution uniqueness. A small pitch $\Lambda$ is therefore instrumental to suppressing aliasing and increasing FOV, a unique advantage of metasurfaces over their classical diffractive counterpart. Another benefit of a small pitch is that the wavefront can be more precisely sculpted at a deep sub-wavelength scale, thereby enhancing optical efficiency[29]. This is particularly important for wide-FOV beam steering at large angles, where the wavefront follows a rapid spatial variation.

Besides aliasing suppression and FOV, the main performance requirements for beam steering devices include beam divergence (which relates to angular resolution), speed, and reliability. Beam divergence is specified by the optical aperture size, the number of meta-atoms and hence complexity of active control, while speed and reliability relate to the metasurface tuning mechanism. For automotive applications, an angular resolution of 0.2° or better is necessary, which translates to an aperture size of ~ 200 μm or larger for near-IR LiDAR. Automotive applications further specify a baseline frame rate of 10 Hz (or higher), a typical combined (horizontal) FOV of 120°, and compliance with reliability standards set forth by IEC, ISO, ASTM and individual auto manufacturers, which mandate tolerance against temperature excursions, mechanical shock, vibrations, fatigue, dust, and salt mist. The reduced temperature sensitivity (as compared to traditional refractive optics) and structural ruggedness of metasurfaces present an additional edge.

Several active metasurface prototypes for beam steering have been demonstrated. Li *et al.* demonstrated a 1-D phase-only SLM based on a LC-infiltrated $TiO_2$ Huygens metasurface with an aperture size of 120 μm × 100 μm[30]. Each electrically addressed pixel comprise three rows of meta-atoms with a combined width of ~ 1 μm. The device achieves a deflection efficiency of 36% at 660 nm wavelength and a FOV of 22°. In parallel, Lumotive, a start-up focusing on LiDAR technologies, has been developing a "Meta-LiDAR" platform based on LCoS[31]. While details of the technology are not available in the public domain, their patents describe LC-infiltrated metal antenna arrays as the active beam steering element.

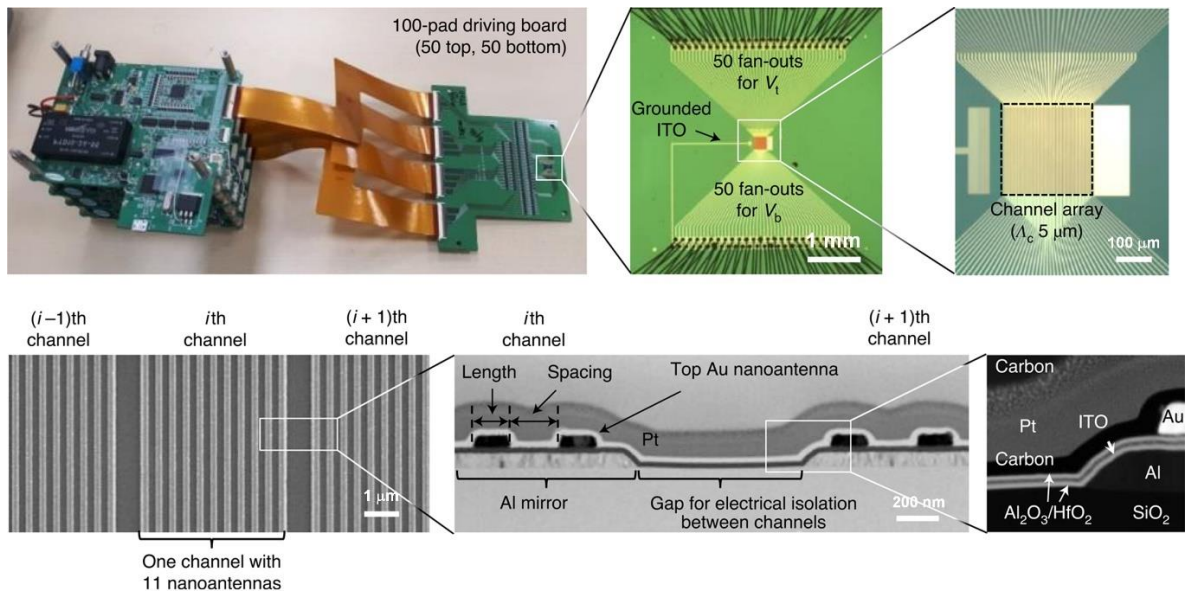

**(Top left) Active metasurface array mounted on an electronics board capable of independently driving 50 channels (pixels) with a pitch of $\Lambda_c$ = 5 μm; (top right) optical microscopy images of the array; (bottom left) scanning electron microscopy image showing the 1-D pixels, each consisting of 11 meta-grating lines; (bottom right) transmission electron microscopy images showing cross-sectional structures of the active array**[5].

Beam steering at higher speed warrants alternative mechanisms. As illustrated in the figure, researchers from Samsung demonstrated active 1-D meta-gratings made of indium tin oxide (ITO), where each individually contacted pixel contains 11 grating lines[5]. A dual gate configuration was employed to realize independent control of phase and amplitude, a useful feature enabling apodization and sidelobe suppression. An RC-limited 3-dB bandwidth of 170 kHz was attained in a 200 μm × 200 μm device with a FOV of 15.4° and a deflection efficiency between 34% to 48%. 2-D beam steering based on a similar mechanism was also reported recently[32].

Before these pioneering demonstrations can enter the commercial realm, considerable performance improvements are anticipated. Down scaling the pixel size to the single meta-atom level will fully leverage the promised advantages of metasurface OPAs such as aliasing-free operation, high efficiency and large FOV. Angle-dependent and nonlocal metasurface designs will further enhance the performance of such large-angle meta-optical systems. A scalable electrical addressing scheme commensurate with large aperture active tuning is sought after to enhance resolution and facilitate agile 2-D beam steering. Finally, reliable packaging suitable for field deployment must be developed and validated.

### Box 3. Active metasurface enabled glasses-free 3-D display

Glasses-free 3-D display (or autostereoscopy) is a technology poised to reshape human-machine interactions. Unlike conventional display panels which reproduces only the intensity of light emanating from an object, an autostereoscopic display restores the light field information including both intensity and propagation direction. The schematic configuration of a 3-D light field display is depicted in the figure. A pixel modulates the light output intensity and at the same time directs emitted light in a specific direction. Therefore, the pixels can be grouped according to their emission directions such that each subset of pixels project a unique perspective view of the displayed scene along one viewing angle (and hence the name "multiview display"), thereby creating 3-D stereoscopic perception for users. Early prototypes of multiview display were implemented using parallax barriers, lenticular lenses or micro-lens arrays on top of flat display panels, although they face limitations in efficiency, FOV and depth of field. Moreover, spatial and angular resolutions represent an inherent trade-off, since the spatial resolution is reduced by a factor equaling to the number of angular views. The subpar spatial and angular resolution is a primary factor that degrades user experience.

The challenges encountered by traditional optics have fueled a growing interest in multiview displays based on flat optics. In 2013, a group from HP Laboratories reported a 3-D display technology using arrays of diffractive grating pixels in place of refractive optics[33], providing 64-view images within 90° FOV. Notwithstanding the passive nature of the grating pixels, they were integrated with an active LC shutter plane to perform dynamic image display. This pioneering innovation has been successfully commercialized by an HP spin-off Leia Inc.[34]. Using metasurfaces as the light-directing pixels promises several additional benefits. While the efficiency of traditional diffraction gratings is limited by power dissipation into high-order diffraction, metasurfaces avoid undesirable diffraction orders to significantly boost efficiency and diminish background noise. The exceptional light bending capability of metasurfaces affords a large FOV without compromising efficiency. For example, recent work by Hua *et al.* demonstrated a metasurface-enabled full-color 3-D display prototype with a record 160° horizontal FOV[35]. Metasurfaces also allow densely packed pixels with a large fill factor to improve display resolution without inducing excessive crosstalk, and they can further exploit temporal and polarization multiplexing schemes to alleviate the trade-off between spatial and angular resolution[36]. Additionally, a metasurface pixel arrays with tunable light directing properties can be coupled with an eye tracking sensor such that the number of angular views broadcasted toward the observer is dynamically optimized. Yet another potential advantage offered by metasurfaces is the prospect of their monolithic integration on display pixels such as micro-LEDs, which opens up unprecedented latitude for fine control of the emission light field[37,38].

In sum, scalable manufacturing and packaging of high-density, large-area metasurface pixel arrays remain the key enabling technology to resolve the resolution bottleneck of 3-D displays and catalyze their widespread adoption in next-generation consumer electronic devices.

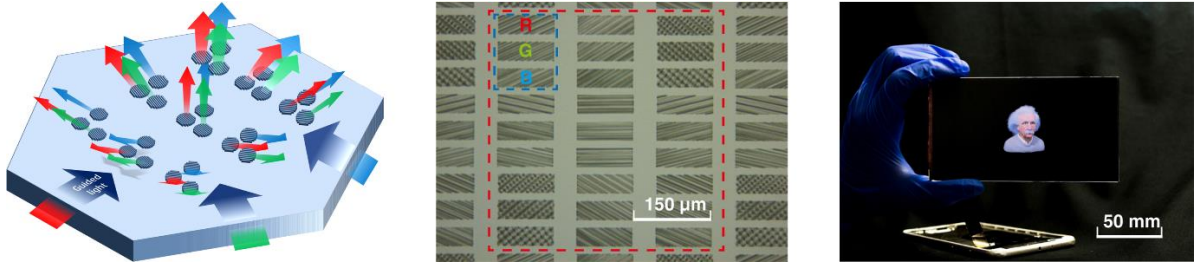

(Left) Schematic design of a multi-view 3-D display panel consisting of metasurface pixel arrays[33]; (middle) optical microscopy image of tri-color metasurface pixel arrays[35]; (right) a full-color, video-rate 3-D display prototype[35].

## Missing links: bridging the present and the prospect

Despite the explosive growth of active metasurface technologies over the past few years, several technological gaps still loom. Here we focus on three critical areas where more R&D efforts are mandated before the application prospects portrayed in the last section can be fulfilled.

### *Scalable manufacturing and packaging*

The first and foremost barrier lies in scalable manufacturing and packaging of active metasurfaces. Their passive counterparts used to encounter the same challenge: in the early phase of development, metasurfaces were almost exclusively prototyped in university cleanrooms using electron beam lithography with painstakingly low throughput. Recent advances have nonetheless circumvented this bottleneck, as fabrication of passive metasurfaces on full glass wafers have been validated via deep ultraviolet (DUV) lithography in standard silicon foundries[50,51] and other high-throughput fabrication methods such as nanoimprint lithography are being explored as well. Manufacturing of active metasurfaces not only stipulates similar requirements on large-area, fine-line lithographic patterning, but is also appreciably more complicated than the passive counterpart. Therefore, commercially viable manufacturing practices of active metasurfaces must maximally leverage standard semiconductor foundry processing and packaging to manage the escalating fabrication and assembly complexity. The foundry manufacturing process may be complemented with backend integration to introduce new materials and functions, in which case the integration process shall capitalize on a mature industrial ecosystem (as summarized in Table 1) to access existing infrastructures, knowledge base, and supply chain to expedite the technology's learning curve.

### *Electrical addressing of large 2-D pixel arrays*

Scaling the existing active metasurface technologies to electrically controlled, high-density and large-size 2-D pixel arrays marks another technical milestone. This is motivated by the demand for enhanced wavefront control with fine resolution: taking beam steering (Box 2) as an example, reducing pixel size contributes to expanding FOV and suppressing sidelobes. At the limit when each meta-atom can be individually tuned (i.e. one meta-atom per pixel), a 'universal' optic results, enabling not only continuous tuning but also reconfiguration of the metasurface to emulate arbitrary optics.

The Holy Grail of a 'universal optic', however, face major challenges in electrical wiring, cross-talk, and control complexity. A small (10 × 10 pixels) 2-D active metasurface array was demonstrated recently[32], although the in-plane fan-out wiring design is not scalable to large arrays. A practical solution for large 2-D arrays involves integrating the pixels to a control backplane via vertical interconnect accesses to form a cross-bar matrix. In the case of volatile pixels, each pixel needs to be coupled with a transistor such that its can be individually tuned, analogous to the active matrix architecture in flat panel displays. For nonvolatile active metasurfaces, a simplified 'passive matrix' configuration without the transistor backplane is equally viable since the 'set-and-forget' pixels can be reconfigured sequentially row-by-row, albeit at the expense of refreshing rate. Notably, each pixel in a passive-matrix cross-bar array still requires a selector with nonlinear I-V characteristics to prevent sneak-path current, which can be implemented via semiconductor diodes or threshold switching phenomena in various materials.

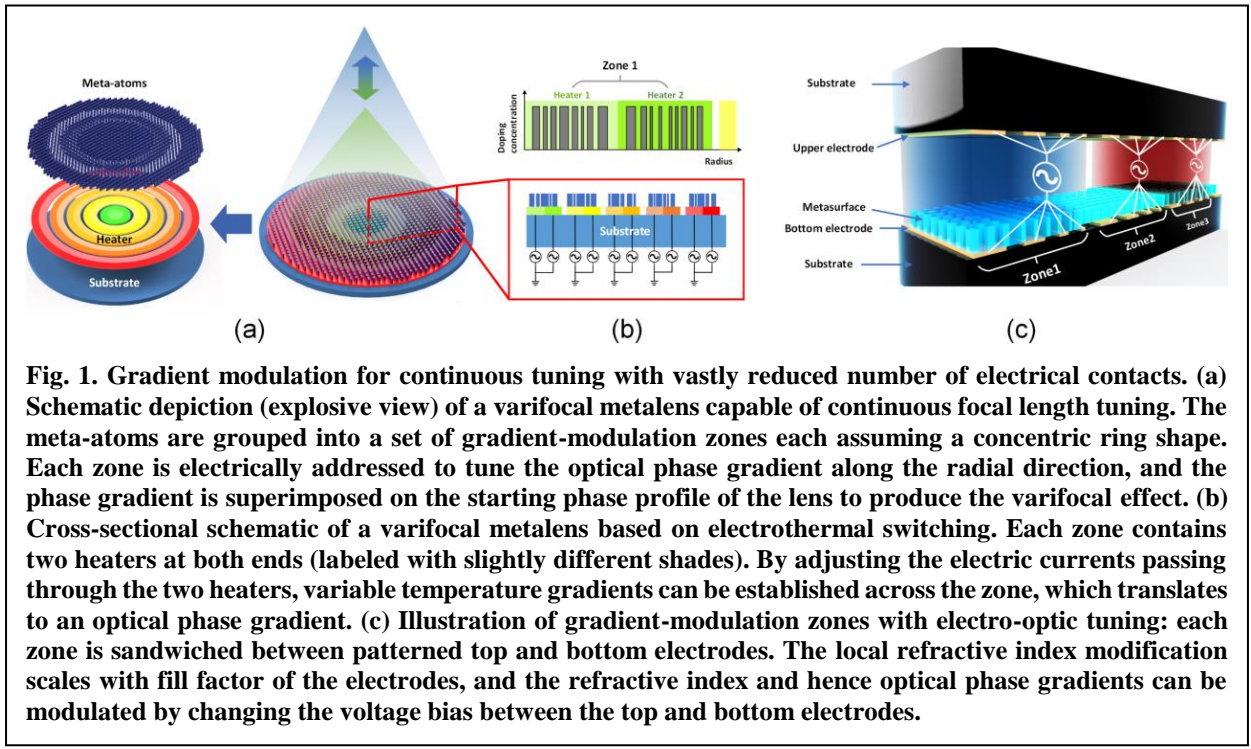

**Fig. 1. Gradient modulation for continuous tuning with vastly reduced number of electrical contacts. (a) Schematic depiction (explosive view) of a varifocal metalens capable of continuous focal length tuning. The meta-atoms are grouped into a set of gradient-modulation zones each assuming a concentric ring shape. Each zone is electrically addressed to tune the optical phase gradient along the radial direction, and the phase gradient is superimposed on the starting phase profile of the lens to produce the varifocal effect. (b) Cross-sectional schematic of a varifocal metalens based on electrothermal switching. Each zone contains two heaters at both ends (labeled with slightly different shades). By adjusting the electric currents passing through the two heaters, variable temperature gradients can be established across the zone, which translates to an optical phase gradient. (c) Illustration of gradient-modulation zones with electro-optic tuning: each zone is sandwiched between patterned top and bottom electrodes. The local refractive index modification scales with fill factor of the electrodes, and the refractive index and hence optical phase gradients can be modulated by changing the voltage bias between the top and bottom electrodes.**

It is worth mentioning that continuous tuning of a meta-optical element does not necessarily entail full 2-D matrix addressing. Here we propose a new concept of gradient modulation: instead of directly tuning the phase delay of meta-atom pixels, electrical contact pairs are used to generate an in-plane refractive index gradient in between each pair, which translates to a tunable phase gradient superimposed on the unmodulated metasurface phase profile. Figures 1b and 1c illustrate the electrode arrangement to produce the variable phase gradient in active metasurfaces based on electrothermal and electro-optic tuning mechanisms, respectively. A functional meta-optical device is assembled from a multitude of gradient-modulation zones defined by the electrode pairs. For instance, Fig. 1a depicts a varifocal metalens comprising zones of concentric ring shapes. The scalable design can dramatically reduce the number of electrical leads and hence wiring complexity compared to 2-D matrix addressing. As an example, we show that a metalens with 1 mm diameter partitioned into merely 23 gradient-modulation zones is capable of continuous focal length tuning from 2 mm to infinity at 2.2 μm wavelength while maintaining diffraction-limited performance throughout (refer to Supplementary Information for more details). In contrast, 2-D addressing of a metalens of the same size involve a staggering half million active pixels!

*Reliability is the key*

Last but certainly not least comes reliability, a topic rarely deliberated in academic publications. Nevertheless, its importance to practical applications cannot be over-emphasized. The reliability requirement is illustrated in the example of automotive LiDAR (Box 2), whose durability qualifications under various environmental stresses are elaborately enumerated by international standards and manufacturers' specifications. For active metasurfaces, endurance is another critical parameter. As is evident from Tables 1 and 2, the endurances of many active metasurface technologies are still lagging application demands. Even for those labeled with 'very large' endurance, their reliability under realistic deployment conditions has often not been verified. On the bright side, unlike their passive counterparts whose performance is fixed post-fabrication, active metasurfaces allow reconfiguration or re-training of their design parameters to "adapt" to imperfections during the fabrication and operation processes, thereby significantly improving system robustness to maintain design goals.

Addressing the reliability challenge warrants scrupulous characterization of degradation kinetics in application-relevant environments, thorough material investigations to elucidate the pertinent failure mechanisms, and judicious device designs guided by the fundamental insights to improve robustness. The studies will inform the essential path for active metasurfaces to make a lasting impact on photonic applications.

**New capabilities: beyond today's functional repertoire**

In addition to filling the missing links, new research initiatives in the field are set to significantly expand the functionalities of active metasurface optics. In this section, we highlight two salient examples where such advances are anticipated to enhance their performances and open up novel application venues.

*Dual modulation of phase and amplitude*

Even though imparting a phase delay without incurring optical loss is a common design prescription for phase-gradient metasurfaces, there are scenarios where simultaneous phase and amplitude modulation comes in handy. In OPAs (Box 2), amplitude apodization helps to match the output field profile to that of a Gaussian beam to suppress sidelobes. In 3-D display (Box 3), the dual modulation is the prerequisite for complete light field manipulation. In analog optical computing, both phase and amplitude are often concurrently used to encode information[1].

A straightforward solution is to assign the amplitude and phase modulation functions to two cascaded active metasurfaces at the expense of fabrication complexity and optical efficiency. Alternatively, a dual-gate configuration has been applied to ITO metasurfaces to accommodate complex reflectance tuning by adjusting two gate voltages[5,39], although the coupled refractive index and absorption changes due to free carrier dispersion in ITO restricts the accessible tuning range. Resorting to two distinct mechanisms to separately engineer refractive index and loss furnishes far more versatile phase and amplitude control. Candidate materials include ionic conductors where migration of different ions produces tri-state switching with orthogonal optical property changes[40], and chalcogenide PCMs where crystallization and vacancy ordering account for decoupled refractive index and absorption modifications[41].

*Metasurface-augmented active photonics*

While useful as standalone optical elements, the application scope of active metasurfaces can be significantly broadened once they can be seamlessly integrated with traditional optical or optoelectronic components. Such metasurface-augmented photonics transcends intrinsic performance limits confronting metasurfaces, such as narrow spectral bandwidth and low quality factor (Q): combining metasurfaces with refractive or reflective optics overcomes their group delay limitation[42] to support broadband operation[43], and embedding active meta-atoms inside a Fabry-Perot cavity maintains the otherwise wavelength-sensitive high-Q resonance condition across a wide tuning range. Besides the performance gains, metasurfaces further profit from their compatibility with monolithic or hybrid integration on optoelectronic platforms (e.g., lasers, detectors, planar photonic integrated circuits, as well as display and imaging arrays) compared to conventional refractive or diffractive elements. The integration can potentially leverage scalable semiconductor fabrication routes to access sophisticated electronic backplanes for massive array modulation. Moreover, some tuning operations (in particular amplitude modulation) can be offloaded to the optoelectronic components to simplify the meta-optical system while retaining active functionalities (e.g., in a metasurface-integrated micro-LED multi-view display alluded to in Box 3). We envision that the metasurface-augmented optoelectronic systems will confer novel capabilities such as pixel-level light field control and detection, computational imaging, high-resolution imaging spectroscopy, and neuromorphic computing.

Realizing the metasurface-augmented active photonics calls for innovative design and fabrication strategies. Specifically, two modeling frameworks are essential to metasurface-augmented photonics. One is a computationally efficient objective-driven method to approach this inherently multiscale – spanning six orders of magnitude in spatial dimensions – and multi-objective – as specified by the active tuning condition – design problem. A successful recipe will likely bridge classical ray based optimization of traditional optics and emerging full-wave inverse design techniques for sub-wavelength meta-structures[44–46]. The other involves multi-physics *predictive* models linking the (electrical, thermal, mechanical, and/or optical) stimuli to the resulting active metasurface response, which becomes particularly important in treating complex material transformations where simple effective medium theory fails[47,48].

Advancing metasurface-augmented photonics also demands fabrication schemes commensurate with scalable manufacturing, such as conformal processing of active metasurfaces on both flat and curved surfaces of conventional optics, and monolithic integration routes of meta-optics on optoelectronic devices[49]. We foresee that breakthroughs in scalable metasurface integration will catalyze new applications exploiting the best of two worlds in both conventional optics and metasurfaces.

**Summary and outlook**

Since their introduction in the past decade, active metasurfaces have swiftly taken the center stage in metamaterials research, boasting significantly enhanced and expanded functionalities over their passive counterparts. The extra layer of complexity needed for active tuning, however, presents an additional barrier to their deployment in the commercial domain.

To overcome the challenges, the imminent success of passive metasurfaces, whose practical applications start to surface, sets a paradigm. Over the past few years, the community have converged upon several potential beachhead markets of passive metasurfaces where their key competitive advantages – system-level SWaP benefits, minimal monochromatic aberration, polarization discrimination capacity, and low-cost at scale – are fully mobilized, and forged a path toward large-area, cost-effective manufacturing capitalizing on standard foundry processing. Likewise, leveraging existing semiconductor foundry infrastructures as well as mature ecosystems in adjacent industries provide a shortcut to facilitate scalable manufacturing and packaging of active metasurface devices. As active metasurfaces establish their manufacturing scalability and reliability, industrial deployment of active metasurfaces will also be initiated first in niche applications, with several promising early examples discussed in the text, before percolating into established markets. In this process, new and unique functionalities exemplified by 'universal optics' consisting of 2-D active meta-atom pixel arrays, electronics and optoelectronics with monolithically integrated metasurface optics, and flat optics rendering complete active control of light phase and amplitude will continue to extend their application venues. The growing market demands will in turn drive the assimilation of active metasurface fabrication into mainstream foundry processes to further enhance yield, lower cost, and improve reliability toward widespread adoption of the technology. This is a bright prospect that the entire active metasurface community can and should strive for, as 'the future depends on what we do in the present'.

**Acknowledgments**

This work was sponsored by the National Science Foundation under award number 2132929, Defense Advanced Research Projects Agency Defense Sciences Office Program: EXTREME Optics and Imaging (EXTREME) under agreement number HR00111720029, the National Institute of Aerospace, and Lockheed Martin Corporation Internal Research and Development. The authors would like to thank Sensong An and Fan Yang for assistance with optical modeling and Xiaochen Sun, Xu Fang, and Lei Bi for helpful technical discussions. The views, opinions and/or findings expressed are those of the authors and should not be interpreted as representing the official views or policies of the Department of Defense or the U.S. Government.

**Author contributions**

All authors contributed to writing the paper.

**Competing financial interests**

The authors declare no competing financial interests.

Table 1. Summary of active metasurface technologies (refer to Supplementary Information for more details and discussions)

| Type | Material or mechanism | Refractive index tuning range | Optical absorption | Endurance (cycling lifetime) | Failure mechanism limiting cycling lifetime | 3-dB bandwidth or 10-90 rise/fall time | Speed limiting factor | Volatility | Foundry manufacturing compatibility | Potential challenges | Relevant industry ecosystem |
|---|---|---|---|---|---|---|---|---|---|---|---|
| Mechanical | Displacement | - | None | Very large | Mechanical failure | ~ 1 Hz | Speed of mechanical motion | - | Compatible | Integration challenge | - |
| | Elastic deformation | - | None | 15,000 | Material fatigue or interface delamination | ~ 1 Hz | Speed of mechanical motion | - | - | Reproducibility and stability | - |
| | MEMS actuation | - | None | > $10^9$ | Material fatigue; stiction; delamination; wear | 1 MHz | Resonant frequencies of mechanical eigenmodes | Design-dependent | Compatible | High voltage | MEMS |
| Free carrier density modulation (electrical injection) | Semiconductors (junction biasing) | 0.08 | Free carrier absorption (FCA) | Very large | - | 0.18 MHz | Carrier transit time; RC delay | Volatile | Compatible with III-V foundry processes | Optical loss; small index change or optical modal overlap with the active region | Semiconductor |
| | TCOs (field gating) | 1.39 | FCA | Very large | - | 10 MHz | | Volatile | Backend processing available in selected foundries | | Display |
| | 2-D materials (field gating) | ~ 1 | Material absorption | Very large | - | > 1 GHz | | Volatile (capacitive) | Backend integration currently under development | | - |
| Thermo-optic | Semiconductors (thermal free-carrier refraction) | 1.5 | FCA | Likely large | - | 5 kHz | Thermal time constant | Volatile | Compatible if only CMOS materials are used | Optical loss; relatively slow response | Integrated photonics |
| | Semiconductors or dielectric materials | 0.15 | Minor FCA due to carrier thermalization | | - | | | | | Relatively slow response; large energy consumption | |
| Electro-optic | EO polymers | 0.11 | None | Very large | - | 50 MHz | RC delay | Volatile (capacitive) | - | High voltage | - |
| | EO crystals | 0.001 | None | Very large | - | 95 MHz | RC delay | Volatile (capacitive) | Backend processing available in selected foundries | Small modulation amplitude | Integrated photonics |
| | Liquid crystals | 0.15 | None | > $10^{10}$ | - | 350 Hz | Relaxation time of liquid crystal molecules | Volatile (capacitive) | Backend processing and packaging available | Relatively slow response | Display |
| | Semiconductor multi-quantum-well (quantum confined Stark effect) | ~ 0.01 | Electroabsorption dictated by the Kramers-Kronig relations | Very large | - | < 10 ns | RC delay | Volatile (capacitive) | Compatible with III-V foundry processes | Full $2\pi$ phase coverage | III-V optoelectronics |
| Phase transition | $VO_2$ | Up to ~ 0.5 in visible and near-IR | FCA in the metallic state | > 24,000 | Defect migration | 450 fs / 2 ps | Kinetics of Mott transition or structural phase transition | Volatile | Backend processing available in selected foundries | Optical loss | - |
| | Chalcogenide PCMs | 3.3 ($Ge_2Sb_2Te_5$) 1.8 ($Ge_2Sb_2Se_4Te$) | None | > $5 \times 10^5$ | Elemental segregation; cyclic strain | 200 ns / 300 ns | Crystallization speed | Non-volatile | Backend processing available in selected foundries | High voltage | Memory |
| Electrochemical | Electrochromic polymers | 0.7 | Electronic absorption at the oxidized state | > $10^7$ | Photochemical degradation | > 25 Hz | Ion transport kinetics in the polymer and electrolyte | Non-volatile | - | Relatively slow response; optical loss | Smart windows |
| | Ionic conducting oxides (protonation) | 0.45 ($SmNiO_3$) ~ 0.4 ($GdO_x$) | FCA in the metallic state | Several hundred | Dielectric breakdown; defect migration | 13 ms | Ion transport kinetics; film thickness | Non-volatile | - | Endurance; relatively slow response | - |
| | Ionic conducting oxides (lithium intercalation) | ~ 0.2 ($WO_3$) 0.65 ($TiO_2$) | FCA in the metallic state | 400 | Ion trapping | 3 s | Ion transport kinetics; film thickness | Non-volatile | - | Endurance; slow response | Smart windows |
| | Metal electrodeposition | - | Absorption of electrodeposited metal | > 200 | Metal oxidation | ~ 1 Hz | Electrodeposition reaction rate | Non-volatile | - | Integration challenge | - |
| Chemical | Metals (hydrogenation) | ~ 4 (Mg) | FCA | 3,000 | Strain due to volume change; oxidation and hydroxylation | ~ 0.1 Hz | Hydrogen diffusion in metal; film thickness | Non-volatile | - | Integration challenge | - |
| | Cover material addition/removal | ~ 0.5 | None | > 10 | Residue and contamination | Seconds to minutes | Speed of dissolution or removal process | Non-volatile | - | Integration challenge | - |
| Magnetic | Magnetooptical oxides | ~ 0.01 | None | Very large | - | 5 GHz | Kinetics of magnetic domain wall motion | Both | - | Integration challenge | - |
| All-optical | Kerr nonlinearity | Light intensity dependent | Nonlinear absorption | Very large | - | 40 fs | Ultrafast electronic nonlinearity | Volatile | Compatible if CMOS materials are used | Integration challenge | - |
| | Free carrier injection | 0.14 | FCA | Very large | - | ~ 100 fs / 20 ps | Carrier lifetime | Volatile | Compatible if CMOS materials are used | Integration challenge | - |

**Table 2. Potential applications of active metasurfaces: green-yellow-gray colors indicate decreasing relevance of the metric to the target use case: very important-somewhat important-minimally relevant (refer to Supplementary Information for more details and discussions)**

| Application | Tuning scheme | Optical tuning parameter (phase/amplitude) | Optical contrast (relevant metrics) | Optical loss suppression | Endurance (cycling lifetime requirement) | Speed (bandwidth requirement) | Power consumption |
|---|---|---|---|---|---|---|---|
| Tunable filters for multispectral sensing | Continuous | Amplitude | Extinction ratio | | $10^7$ | 1 kHz | |
| Beam steering for LiDAR | Continuous | Both | Full $2\pi$ phase tuning range | | $10^9$ | 10 Hz | |
| Light field display | Continuous | Both | Field-of-view and image contrast | | $10^{10}$ | 30 Hz | |
| Computational imaging | Discrete | Phase | Full $2\pi$ phase tuning range | | $10^{10}$ | 100 Hz | |
| Optical neural network with adaptive network training | Continuous | Both | Full $2\pi$ phase tuning range | | $10^8$ | 1 kHz | |
| Dynamic projection display | Continuous | Amplitude | Image contrast | | $10^{10}$ | 30 Hz | |
| Electronic paper (reflective display) | Discrete or continuous | Amplitude | Color saturation and image contrast | | $10^7$ | 1 Hz | Nonvolatile or capacitive |
| Zoom lens | Discrete or continuous | Phase | Full $2\pi$ phase tuning range | | $10^5$ | 1 Hz | |
| Digital signal modulation for free-space communications | Discrete | Either | Modulation contrast | | $10^{18}$ | 10 GHz | |
| Adaptive optics | Continuous | Phase | Full $2\pi$ phase tuning range | | $10^{10}$ | 100 Hz | |
| Nonreciprocal optics based on spatiotemporal modulation | Discrete | Either | Isolation ratio | | $10^{18}$ | 10 GHz | |
| Optical limiter | Discrete | Amplitude | Extinction ratio | | Application-specific | > 1 GHz | Nonvolatile |
| Adaptive thermal camouflage | Continuous | Amplitude | Dynamic range | | $10^8$ | 10 Hz | |

# Active metasurfaces: lighting the path to commercial success

Tian Gu[1,2,*], Hyun Jung Kim[3,4,*], Clara Rivero-Baleine[5,*], and Juejun Hu[1,2,*]

[1]Department of Materials Science & Engineering, Massachusetts Institute of Technology, Cambridge, Massachusetts, USA
[2]Materials Research Laboratory, Massachusetts Institute of Technology, Cambridge, Massachusetts, USA
[3]National Institute of Aerospace, Hampton, Virginia, USA
[4]NASA Langley Research Center, Hampton, Virginia, USA
[5]Missiles and Fire Control, Lockheed Martin Corporation, Orlando, Florida, USA

*gutian@mit.edu, hyunjung.kim@nasa.gov, clara.rivero-baleine@lmco.com, hujuejun@mit.edu

**Table S1. Extended summary of active metasurface technologies**[a]

| Type | Material or mechanism | Refractive index tuning range | Optical absorption | Endurance (cycling lifetime) | Failure mechanism limiting cycling lifetime | 3-dB bandwidth or 10-90 rise/fall time[b] | Speed limiting factor | Volatility | Foundry manufacturing compatibility | Potential challenges | Relevant industry ecosystem |
|---|---|---|---|---|---|---|---|---|---|---|---|
| Mechanical | Displacement | - | None[c] | Very large | Mechanical failure | ~ 1 Hz | Speed of mechanical motion | - | Compatible | Integration challenge | - |
| | Elastic deformation | - | None[c] | 15,000 [1] | Material fatigue or interface delamination | ~ 1 Hz | Speed of mechanical motion | - | - | Reproducibility and stability | - |
| | MEMS actuation | - | None[c] | > $10^9$ [2] | Material fatigue; stiction; delamination; wear | 1 MHz [2] | Resonant frequencies of mechanical eigenmodes | Design-dependent | Compatible | High voltage | MEMS |
| Free carrier density modulation (electrical injection) | Semiconductors (junction biasing) | 0.08 [3][d] | Free carrier absorption (FCA) | Very large [4][e] | - | 0.18 MHz [5] | Carrier transit time; RC delay | Volatile | Compatible with III-V foundry processes | Optical loss; small index change or optical modal overlap with the active region | Semiconductor |
| | TCOs (field gating)[f] | 1.39 [6][g] | FCA | Very large | - | 10 MHz [7] | | Volatile | Backend processing available in selected foundries | | Display |
| | 2-D materials (field gating) | ~ 1 [8] | Material absorption | Very large | - | > 1 GHz [9] | | Volatile (capacitive) | Backend integration currently under development [10] | | - |
| Thermo-optic | Semiconductors (thermal free-carrier refraction) | 1.5 [11][g] | FCA | Likely large [12][h] | - | 5 kHz [13] | Thermal time constant | Volatile | Compatible if only CMOS materials are used | Optical loss; relatively slow response | Integrated photonics |
| | Semiconductors or dielectric materials | 0.15 [14][i] | Minor FCA due to carrier thermalization | | | | | | | Relatively slow response; large energy consumption | |
| Electro-optic | EO polymers | 0.11 [15][j] | None[c] | Very large | - | 50 MHz [15,16] | RC delay | Volatile (capacitive) | Backend processing available in selected foundries | High voltage | - |
| | EO crystals | 0.001 [17] | None[c] | Very large | - | 95 MHz [18] | RC delay | Volatile (capacitive) | Backend processing available in selected foundries [19,20] | Small modulation amplitude | Integrated photonics |
| | Liquid crystals | 0.15 [21,22][k] | None[c] | > $10^{10}$ [l] | - | 350 Hz [23] | Relaxation time of liquid crystal molecules | Volatile (capacitive) | Backend processing and packaging available | Relatively slow response | Display |
| | Semiconductor multi-quantum-well (quantum confined Stark effect) | ~ 0.01 [24] | Electroabsorption dictated by the Kramers-Kronig relations | Very large | - | < 10 ns [25] | RC delay | Volatile (capacitive) | Compatible with III-V foundry processes | Full 2π phase coverage | III-V optoelectronics |
| Phase transition | $VO_2$ | Up to ~ 0.5 in visible and near-IR [26,27] | FCA in the metallic state | > 24,000 [28][m] | Defect migration [29] | 450 fs / 2 ps [30] † | Kinetics of Mott transition or structural phase transition[n] | Volatile[o] | Backend processing available in selected foundries | Optical loss | - |
| | Chalcogenide PCMs | 3.3 ($Ge_2Sb_2Te_5$) [31][p] 1.8 ($Ge_2Sb_2Se_4Te$) [32][p] | None [32][c] | > 5 × $10^5$ [33][q] | Elemental segregation; cyclic strain [34] | 200 ns / 300 ns [35] † | Crystallization speed [36] | Non-volatile[r] | Backend processing available in selected foundries [37,38] | High voltage | Memory |
| Electrochemical | Electrochromic polymers | 0.7 [39][s] | Electronic absorption at the oxidized state | > $10^7$ [40] | Photochemical degradation [41] | > 25 Hz [40][t] | Ion transport kinetics in the polymer and electrolyte [42] | Non-volatile | - | Relatively slow response; optical loss | Smart windows |
| | Ionic conducting oxides (protonation) | 0.45 ($SmNiO_3$) [43][p] ~ 0.4 ($GdO_x$) [44][u] | FCA in the metallic state | Several hundred [44][v] | Dielectric breakdown [44]; defect migration | 13 ms [44] | Ion transport kinetics; film thickness | Non-volatile | - | Endurance; relatively slow response | - |
| | Ionic conducting oxides (lithium intercalation) | ~ 0.2 ($WO_3$) [45][u] 0.65 ($TiO_2$) [46][w] | FCA in the metallic state | 400 [46] | Ion trapping [47] | 3 s [48] | Ion transport kinetics; film thickness | Non-volatile | - | Endurance; slow response | Smart windows |
| | Metal electrodeposition | - | Absorption of electrodeposited metal | > 200 [49][x] | Metal oxidation | ~ 1 Hz [49] | Electrodeposition reaction rate | Non-volatile | - | Integration challenge | - |
| Chemical | Metals (hydrogenation) | ~ 4 (Mg) [50][p] | FCA | 3,000 [51] | Strain due to volume change; oxidation and hydroxylation | ~ 0.1 Hz [52] | Hydrogen diffusion in metal; film thickness [53] | Non-volatile | - | Integration challenge | - |
| | Cover material addition/removal | ~ 0.5 [54,55] | None[c] | > 10 [56] | Residue and contamination [57] | Seconds to minutes | Speed of dissolution or removal process | Non-volatile | - | Integration challenge | - |
| Magnetic | Magnetooptical oxides | ~ 0.01 [58] | None[c] | Very large | - | 5 GHz [59][y] | Kinetics of magnetic domain wall motion | Both | - | Integration challenge | - |
| All-optical | Kerr nonlinearity | Light intensity dependent | Nonlinear absorption | Very large | - | 40 fs [60] | Ultrafast electronic nonlinearity | Volatile | Compatible if CMOS materials are used | Integration challenge | - |
| | Free carrier injection | 0.14 [61] | FCA | Very large | - | ~ 100 fs / 20 ps [62] † | Carrier lifetime | Volatile | Compatible if CMOS materials are used | Integration challenge | - |

[a] Here we focus on active metasurfaces operating in the optical frequency range, specifically from ultraviolet to long-wave infrared (excluding the terahertz spectrum). Only experimental demonstrations are included in the table.
[b] In the case of active metasurfaces with asymmetric rise/fall responses (marked with †), both rise and fall time values are reported.
[c] Here 'none' implies that the optical loss *can be* negligible throughout the entire active tuning cycle. However, this condition does not necessarily hold for all materials or devices belonging to the category.
[d] Refractive index change of Si at 1550 nm wavelength with $10^{20}$/$cm^3$ hole concentration.
[e] Wear-out failure rates of semiconductor modulator devices have been found to be negligible (below 1 Failures in Time, FIT).
[f] In addition to TCO, the same field gating configuration can also be applied to modulate carrier concentrations in semiconductor materials to realize optical tuning within their optical transparency window, for example in [124].
[g] The refractive index change was amplified by operating near the epsilon-near-zero (ENZ) point.
[h] Thermo-optic devices exhibit minimal degradation after 5,000 hours of accelerated aging tests.
[i] Index change in a Si metasurface when temperature increases from 300 K to 800 K at 1.1 µm wavelength.
[j] Index change quoted at 1500 nm wavelength for an applied electric field varying from -100 to 100 V/µm.
[k] Refractive index change of E7 liquid crystal (from Merck, a formulation commonly used in active metasurfaces) in the near-infrared when transitioning from the nematic to the isotropic phase.
[l] While the lifetime of liquid crystal metasurfaces has not been extensively investigated, commercial liquid crystal displays usually specify a lifetime of up to 60,000 hours, which at 60 Hz refresh rate corresponds to > $10^{10}$ switching cycles.
[m] Over 6.8 × $10^8$ switching cycles have been demonstrated in $VO_2$-based RF switches [125].
[n] Mott-Hubbard type phase transition induced by optical pumping of carriers exhibits ultrafast transition kinetics down to 26 fs [126]. In contrast, structural phase transition in $VO_2$ is slower and occurs at a sub-nanosecond time scale or longer [127].
[o] $VO_2$ devices can operate in a nonvolatile mode within a finite temperature range defined by its phase-transition hysteresis [128].
[p] Refractive index change at 1550 nm wavelength.
[q] Endurance values of over 1.5 × $10^8$ and 2 × $10^{12}$ have been demonstrated in RF switches [35] and electronic phase change memories [129] based on chalcogenide materials, respectively.
[r] Volatility of switching phase change materials can be tuned by adjusting pulse parameters [130].
[s] Refractive index change at 633 nm wavelength.
[t] Switching time down to 100 µs has been observed in polyaniline (PANI) [131].
[u] Refractive index change at 400-800 nm wavelengths.
[v] In memristive electronic device made of the rare earth nickelate $NdNiO_3$, stable performance over 1.6 × $10^6$ cycles have been demonstrated [132].
[w] Refractive index change at 649 nm wavelength associated with transition from $TiO_2$ to $Li_{0.5}TiO_2$.
[x] 5,500 reversible electrodeposition cycles of metals have been realized in large-area smart windows [133].
[y] Here the magnetic state is controlled by ultrafast optical excitation. Current driven magnetic switching with a 3-dB bandwidth of 6 GHz has been achieved in integrated magnetooptical waveguide modulators at cryogenic temperatures [134].

**Table 2. Potential applications of active metasurfaces [63–77]: green-yellow-gray colors indicate decreasing relevance of the metric to the target use case: very important-somewhat important-minimally relevant (refer to Supplementary Information for more details and discussions)**

| Application | Tuning scheme | Optical tuning parameter (phase/amplitude)[a] | Optical contrast (relevant metrics) | Optical loss suppression | Endurance (cycling lifetime requirement) | Speed (bandwidth requirement) | Power consumption |
|---|---|---|---|---|---|---|---|
| Tunable filters for multispectral sensing [78–80] | Continuous | Amplitude | Extinction ratio | | $10^{7\ b}$ | 1 kHz[c] | |
| Beam steering for LiDAR [81,82] | Continuous | Both | Full $2\pi$ phase tuning range[d] | | $10^{9\ e}$ | 10 Hz | |
| Light field display [83–85] | Continuous | Both | Field-of-view and image contrast | | $10^{10\ f}$ | 30 Hz[g] | |
| Computational imaging [86] | Discrete | Phase | Full $2\pi$ phase tuning range | | $10^{10\ ee}$ | 100 Hz[h] | |
| Optical neural network with adaptive network training [87,88] | Continuous | Both | Full $2\pi$ phase tuning range | | $10^{8\ i}$ | 1 kHz | |
| Dynamic projection display [89–91] | Continuous | Amplitude | Image contrast | | $10^{10\ ee}$ | 30 Hz[ff] | |
| Electronic paper (reflective display) [92–94] | Discrete or continuous | Amplitude | Color saturation and image contrast | | $10^{7\ j}$ | 1 Hz | Nonvolatile or capacitive |
| Zoom lens [95] | Discrete or continuous | Phase | Full $2\pi$ phase tuning range | | $10^{5\ k}$ | 1 Hz | |
| Digital signal modulation for free-space communications [15,96,97] | Discrete | Either | Modulation contrast | | $10^{18\ l}$ | 10 GHz | |
| Adaptive optics [98] | Continuous | Phase | Full $2\pi$ phase tuning range | | $10^{10\ m}$ | 100 Hz [99] | |
| Nonreciprocal optics based on spatiotemporal modulation [100–102] | Discrete | Either | Isolation ratio | | $10^{18\ kk}$ | 10 GHz | |
| Optical limiter [103,104] | Discrete | Amplitude | Extinction ratio | | Application-specific | > 1 GHz | Nonvolatile |
| Adaptive thermal camouflage [105,106] | Continuous | Amplitude | Dynamic range | | $10^{8\ n}$ | 10 Hz[o] | |

[a] Note that amplitude modulation can be implemented with phase modulation via interference effects although the reverse is not true: phase modulation cannot be realized by amplitude modulation alone.
[b] Taking the Stratospheric Aerosol and Gas Experiment as an example, the science observation events are a maximum length of 6 minutes, and the filter will be switched 5 times per second during the event. Normal low earth orbits are approximately 90 minutes long, and one science observation event will take place per orbit. Approximately $10^7$ cycles are performed per year on orbit based on the estimation.
[c] Specification of high-speed motorized filter wheels (~ 1 ms per filter).
[d] It is possible to realized beam steering (or generally other optical functions as well) with phase coverage well below $2\pi$ albeit at the expense of optical performance [135–137].
[e] Assume 20-year lifetime with 5 hour daily use at 10 Hz frame rate.
[f] Assume 100,000-hour lifetime (comparable to that of state-of-the-art display modules) at 30 Hz frame rate.
[g] National Television Standards Committee (NTSC) video frame rate.
[h] The speed must be higher than the frame rate of single images.
[i] Assume 10,000 in-situ adaptive training iterations, 10 meta-optics layers, and 100 reconfigurations.
[j] Comparable to the lifetime of E-ink displays.
[k] Typical mechanical lenses are capable of ~ 100,000 cycles.
[l] Assume 5-year lifetime at 10 GHz modulation.
[m] Assume 30,000 hour lifetime at 100 Hz.
[n] Assume 3,000 hour lifetime at 10 Hz.
[o] The speed needs to be at least better than that of low frame rate infrared cameras which is generally specified at 9 Hz.

**Continuous tuning via gradient modulation: managing electrical addressing complexity**

The basic concept of gradient modulation is discussed in the main text. One important aspect of the design is that the optical phase profile of the metasurface device is numerically optimized at each optical state rather than analytically defined *a priori*. In the metalens example, it means that the phase profiles do not assume the classical hyperbolic form; instead, a direct search algorithm was employed to optimize the phase gradient in each zone to maximize the optical intensity at the focal spot without constraining the phase [107–111]. The division of the zones is also not defined analytically *a priori* but instead optimized to maximize the mean Strehl ratio of the focal spots. We note that this approach is also conceptually reminiscent of the 'zone engineering' scheme for achromatic metalens design in that the optimized phase profiles are piece-wise smooth with abrupt changes at the zone boundaries [112].

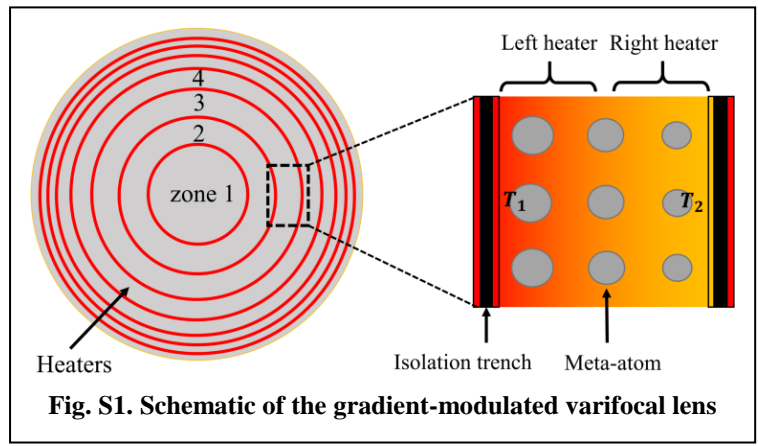

Fig. S1. Schematic of the gradient-modulated varifocal lens

As an example illustrating the gradient modulation design, we consider a PCM-based transmissive active metalens (Fig. S1). The PCM meta-atoms are patterned on a silicon-on-insulator (SOI) substrate, and the SOI layer is selectively doped to create electrically conductive channels in order to electrothermally trigger structural transitions in the PCM meta-atoms. Here the SOI layer acts as an infrared-transparent resistive heater backplane [38,113–115]. Alternatively, other transparent heater materials such as transparent conducting oxides [116–118] and graphene [119,120] can also be used. The doping region is lithographically patterned, which can be implemented using a patterned $SiO_2$ layer as a mask during ion implantation. To generate a temperature gradient across a gradient-modulation zone along the radial direction, we patterned two doped heaters, one at each end of a zone. The widths of the doped regions (with the shape of a set of concentric rings) are chosen such that when the two heaters are separately biased, a nearly uniform temperature gradient is established throughout the entire zone. As an example, Fig. S2a plots the doping profile along the radial direction in a zone, where the doped regions belonging to the two heaters are labeled with different colors. The doped regions are connected in parallel within each heater. By applying different voltages at the two heaters, steady-state temperature profiles between 280 °C and 360 °C with varying gradients can be attained (Fig. S2b). Since we have previously quantified how the refractive index change in a PCM due to structural transition depends on the annealing temperature [36], the thermal gradient translates to locally varying refractive indices of the PCM, in this specific case $Ge_2Sb_2Se_4Te$ or GSST [32,121,122]. The index change then modifies the optical phase delay imparted by PCM meta-atoms, enabling full 2π phase tuning range as the meta-atoms are progressively transformed from amorphous to crystalline state [123]. This mechanism thus allows a variable 'titled' phase profile to be superimposed onto the starting phase distribution in each zone. By optimizing the variable phase gradient in each zone, we show that the focal length of the lens can be tuned to arbitrary values in a pre-defined range (2 mm to infinity in this design example). Figures S3a shows the simulated Strehl ratio of the varifocal metalens as a function of focal length $f$, and Figs. S3b and S3c plot the intensity profiles of the focal spot at $f = 2$ mm and $f = 6$ mm. The results indicate that the lens maintains diffraction-limited performance with a Strehl ratio greater than 0.9 throughout the tuning range, confirming the excellent optical quality of the varifocal metalens following the gradient modulation design.

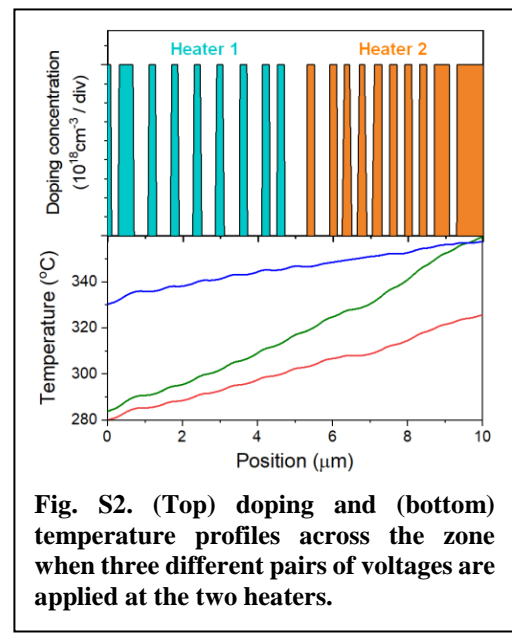

Fig. S2. (Top) doping and (bottom) temperature profiles across the zone when three different pairs of voltages are applied at the two heaters.

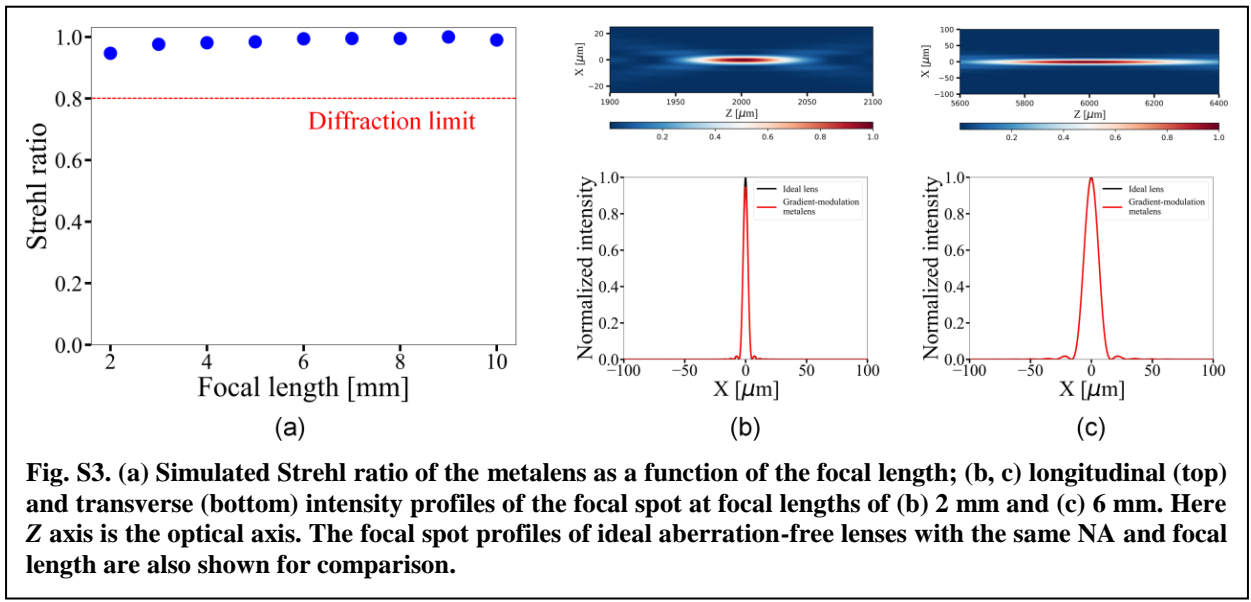

**Fig. S3. (a)** Simulated Strehl ratio of the metalens as a function of the focal length; **(b, c)** longitudinal (top) and transverse (bottom) intensity profiles of the focal spot at focal lengths of **(b)** 2 mm and **(c)** 6 mm. Here Z axis is the optical axis. The focal spot profiles of ideal aberration-free lenses with the same NA and focal length are also shown for comparison.